\documentclass[intlimits,twoside,a4paper]{article}

\usepackage[cp1251]{inputenc}

\usepackage[eqsecnum]{cmpj3}

\usepackage{bm}

\usepackage{amsfonts}
\usepackage{bbold}
\usepackage{mathrsfs}
\usepackage{color}
\usepackage{epstopdf}
\usepackage{graphicx}
\usepackage{bm}
\usepackage{silence}
\usepackage{amssymb,dsfont}

\newcommand{\rr}{  {\mathbf r} }
\newcommand{\ee}{ {\mathbf e} }

\newcommand{\BEQ}{\begin{equation}}
\newcommand{\EEQ}{\end{equation}}
\newcommand{\BEA}{\begin{eqnarray}}
\newcommand{\EEA}{\end{eqnarray}}

\newcommand{\mttp}[1]{\textcolor{black}{#1}}

\issue{2019}{22}{4}{43608}
\doinumber{10.5488/CMP.22.43608}

\begin{document}
\title[Low-frequency excitations and their localization properties in glasses]{
Low-frequency excitations and their localization properties in glasses
}

\author[M. Paoluzzi, L. Angelani ]{M. Paoluzzi\refaddr{adr1}\footnote{Matteo.Paoluzzi@roma1.infn.it}, L. Angelani\refaddr{adr1,adr2}}

\addresses{
\addr{adr1} Dipartimento di Fisica, Sapienza Universit\`a di Roma, Piazzale A. Moro, 2, I-00185 Rome, Italy
\addr{adr2} ISC-CNR,  Institute  for  Complex  Systems,  Piazzale  A.  Moro,  2,  I-00185  Rome,  Italy 
}

\date{Received	May 31, 2019}

\maketitle	

\begin{abstract}
Besides the dynamical slowing down signaled by an enormous increase of the viscosity approaching the glass transition, 
structural glasses show interesting anomalous thermodynamic features at low temperatures that hint at peculiar deviations
from Debye's law at low enough frequencies. 
Theory, numerical simulations, and experiments suggest that deviation from Debye's law is due to soft-localized {\it glassy modes} that populate
the low-frequency spectrum.
We study the localization properties of the
low-frequency modes in a three-dimensional supercooled liquid model. 
The density of states $D(\omega)$ is computed considering the inherent structures of configurations well thermalized at parental temperatures
close to the dynamical transition \mttp{$T_\text d$}.
We observe a crossover in the probability distribution of the
inverse of the participation ratio that happens approaching $T_\text d$ \mttp{from high temperatures}. 
\mttp{We show that a similar crossover is observed at high parental temperature when the translational
invariance of the system is explicitly broken by a random pinning field.}


\pacs 61.20.Lc, 61.20.Ja, 63.50.+x
 \end{abstract}

\section{Introduction}\label{Introduction}

Although glasses and amorphous materials are widespread in nature since the ancient times~\cite{binder2011glassy,leuzzi2007thermodynamics},
a unified and coherent theoretical framework for describing their thermodynamical and dynamical properties remains still a challenge
that attracts the attention of a wide scientific community  \cite{Berthier2011,debenedetti2001supercooled}.
Glasses can be obtained by cooling fast enough a liquid in order to avoid crystallization. 
In experiment and numerical simulations, \mttp{the glass transition temperature is defined as the temperature at
which the structural relaxation time $\tau_\alpha$ overcomes some threshold value.}
A glass can thus \mttp{be} seen as a fluid that does not flow anymore \cite{cavagna2009supercooled}. 
Under this perspective, static observables that \mttp{are usually suitable for revealing} positional order as the radial distribution function $g(r)$ and its Fourier transform, i.e.,
the static structure factor $S(q)$, do not indicate remarkable differences between the liquid and the glassy state.

Looking at glasses with the lens of solid state physics, it turns to be natural to study their low-energy excitations.
In particular, on large scales, glasses are continuum media and thus, at small enough frequencies, 
the density of states $D(\omega)$ follows Debye's law \cite{kittel2005introduction}, i.e., $D(\omega) \sim \omega^{d-1}$ in $d$ spatial dimensions.
Debye’s law assumes that the only low energy excitations are phonons. Debye's law provides precise theoretical predictions for the thermodynamic 
quantities such as the specific heat at low temperatures. However, differently from crystalline solids, glasses show anomalies in thermal
conductivity and specific heat as temperature decreases towards zero \cite{Exp1}. For instance, the specific heat $C_v$ scales with $T$ deviating
from Debye's law. Moreover, thermodynamic anomalies are shared by different glassy systems suggesting that universal mechanisms are responsible 
for that \cite{Exp1}.

Theoretical models as the two-level system model or the soft potential model face the problem looking at other excitations mechanisms
besides phonons that should be taken into account for correctly describing the low excitations in disordered media.
In particular, Gurarie and Chalker in reference \cite{Gurarie03} pointed out that non-Goldstone, and thus non-phononic, excitations in
disordered systems contribute to $D(\omega)$ with a low-frequency sector that is universal, i.e., independent \mttp{of} the spatial dimensions, 
with a scaling $D(\omega)\sim \omega^4$.  Moreover, such {\it glassy modes} are spatially localized and not extended like phonons.
Since in five or less spatial dimensions, the predicted non-Goldstone contribution is subdominant with respect to
Debye spectrum, it is hard to detect in both numerical simulations and experiments. 
Recently, a few numerical strategies have been developed in numerical simulations for taking access to the non-Goldstone
contribution \cite{Baity-Jesi15,Lerner2016,Mizuno14112017,pnas,wang2019low}.

A simple strategy for probing the non-Goldstone sector of the spectrum consists of removing low-frequency phonons. This can be done
introducing an external field that breaks the translational invariance of the system \cite{Baity-Jesi15}. 
Random pinning has been widely adopted in numerical simulations, analytical computations, and experiments to gain an
insight into glassy transition, as a strategy for reaching the Kauzmann temperature and for measuring static correlations lengths \cite{cammarota2013random_pin,cammarota2012ideal_pin,Szamel,kob2012non_pin,gokhale2014growing_pin,karmakar2013random,PhysRevLett.110.245702_pin,ozawa2015equilibrium_pin}.  
In a previous work, we showed
that random pinning can be employed for probing the non-Debye spectrum \cite{pnas}. 

In reference \cite{pnas} we showed that the low-frequency spectrum in a three-dimensional model of glass can be written as $D(\omega) \sim \omega^{s(p)}$,
with $p$ being the fraction of frozen particles. The exponent $s(p)$ turns to be bounded by two extreme values, i.e., $s(p)=\mttp{2}$ for $p \to 0$ and $s(p) = 4$ above 
a threshold value $p_\text{th}$ that is of the order of $50~\%$ of frozen particles.
Such a phenomenology has a simple interpretation: as the number of frozen particles increases, phonons are pushed at higher frequencies and 
the moving particles rattle in a random environment. In particular, their equilibrium positions result to be randomly displaced with respect a crystalline 
configuration and thus these vibrations naturally give rise to a Rayleigh-type scattering mechanism \cite{Exp1}.

Frozen particles provide an artificial tool for introducing heterogeneous regions with different elastic properties. 
We showed that a remarkably similar phenomenology emerges approaching the dynamical transition \cite{arxiv}. 
In particular, as it has been observed in reference \cite{Lerner_rapid}, the low-frequency spectrum of $D(\omega)$ depends
on the parental temperature $T$.
We observed that one can write $D(\omega) \sim \omega^{s(T)}$ with $s(T)=2$ for parental temperatures 
$T \gg T_\text d$, with $T_\text d$ the dynamical temperature, i.e., the temperature where the system undergoes the dynamical arrest.
As $T \to T_\text d$ we observed an increase in $s(T)$, i.e., $s(T) \to 4$ for $T\to T_\text d$.
This happens because of the dynamical heterogeneities \cite{Kob1997} that proliferate as temperature decreases towards the
dynamical temperature. Since the behavior of $s(T)$ mirrors that of $s(p)$, it has been shown that the growing of spatially heterogeneous
regions can be measured comparing the two systems. In this way, one can extract a dynamical correlation length $\xi_\text{pin} \propto p$
as a function of the parental temperature $T$.
$\xi_\text{pin}(T)$ shows a mild divergence at $T_\text d$ which is in agreement with the behavior of the dynamical correlation length $\xi_\text{dyn}$
computed through multi-point correlation functions.

In this paper, we investigate the localization properties of the low-frequency modes of a three-dimensional glass former. 
Measuring the degree of localization of a mode of frequency $\omega$ through the inverse of its participation ratio $\mathcal{R}(\omega)$,
we find that $\mathcal{R}(\omega)$ \mttp{depends on the} parental temperature $T$  for modes populating the low-frequency spectrum.
The low-frequency spectrum is defined as the frequencies below the boson peak that contribute to $D(\omega)$.
In particular, approaching the dynamical transition, the probability distribution function of $\mathcal{R}(\omega)$ shifts towards
higher values. Defining $\mathcal{R}_a$ as the first moment of the distribution, it turns out that $\mathcal{R}_a$ undergoes a smooth crossover mirroring that in $D(\omega) \sim \omega^{s(T)}$.
We then compare the localization properties of low-frequency modes obtained considering a fraction $p$ of particles frozen during the 
minimization of the mechanical energy.  We obtain that also in the case of the pinned system, $\mathcal{R}_{a}$ starts growing 
as $p$ increases. We then compare the two protocols showing that it is possible to extract the behavior of a typical length scale $\xi^3 \propto p$
through the relation $\mathcal{R}_a(T,p=0)=\mathcal{R}_a(T=\infty, p)$. \mttp{The mapping confirms that the properties of the pinned
system at $p\to p_\text{th}$, with $p_\text{th}\sim0.5$ provide complementary information on the same system at $p=0$ and $T\to T_\text d$.
In particular, the dependency $p=p(T)$ is in agreement with other estimates \cite{arxiv}.}

\newpage
\section{Model}
We consider a three-dimensional system composed of  a  $50\!:\!50$ binary
mixture of $N$ soft sphere\mttp{s} confined in a cubic box of side $L$ with periodic 
boundary conditions and interacting  through a pure repulsive pairwise potential \cite{Bernu87,Grigera01}. 
We label large particles with $A$ and the small ones with $B$. The total number of particle\mttp{s} reads $N=N_A+N_B$ and the corresponding density is $\rho=N/L^3$. 
The radii are $\sigma_A$ and $\sigma_B$ with $\sigma_A/\sigma_B\!=\!1.2$ 
and $\sigma_A\!+\!\sigma_B\!\equiv\! \sigma \!=\!1$ \cite{Grigera01}.
The side of the box is $L=\!=\! N^{1/3}$ such that $\rho=1$.
Indicating with $\rr_i$ the position of the particle $i$, with $i=1,\ldots,N$,
two particles $i,j$ interact via the potential  $\phi(r_{ij})=\epsilon [({\sigma_i + \sigma_j })/{r_{ij}}]^{12} + k_0  + k_2 r_{ij}^2\,,$ {\textcolor{black}{where $r_{ij}\!\equiv \!|\rr_i - \rr_j|$}}.
We impose a cutoff to the potential at ${\textcolor{black}{r_\text c\!=\! \sqrt{ 3} \sigma}}$ in a way
that $\phi(r)=0$ for $r>r_\text c$. The coefficients $k_0$ and $k_2$ ensure a continuity to 
$\phi(r)$ up to the first derivative at $r=r_\text c$.

\subsection{Equilibrium dynamics}
\textcolor{black}{For the dynamics, we have considered hybrid Brownian/Swap Monte Carlo simulations obtained combining the numerical integration of the equations of motion with Swap Monte Carlo moves \cite{Grigera01}.}
In particular, \textcolor{black}{in order to generate thermalized configurations}, we propose an update of the system
according to swap moves \textcolor{black}{every $2 \times 10^3$} time steps.
We consider system sizes $N=10^3,12^3$ and averaging over $400$ \mttp{independent} configurations.
In what follows we report all quantities in reduced units \textcolor{black}{considering $\sigma=\epsilon=\mu=1$, where $\mu$ is the mobility
of the Brownian particles}.

\begin{figure}[!b]
\centering
\includegraphics[width=.5\textwidth]{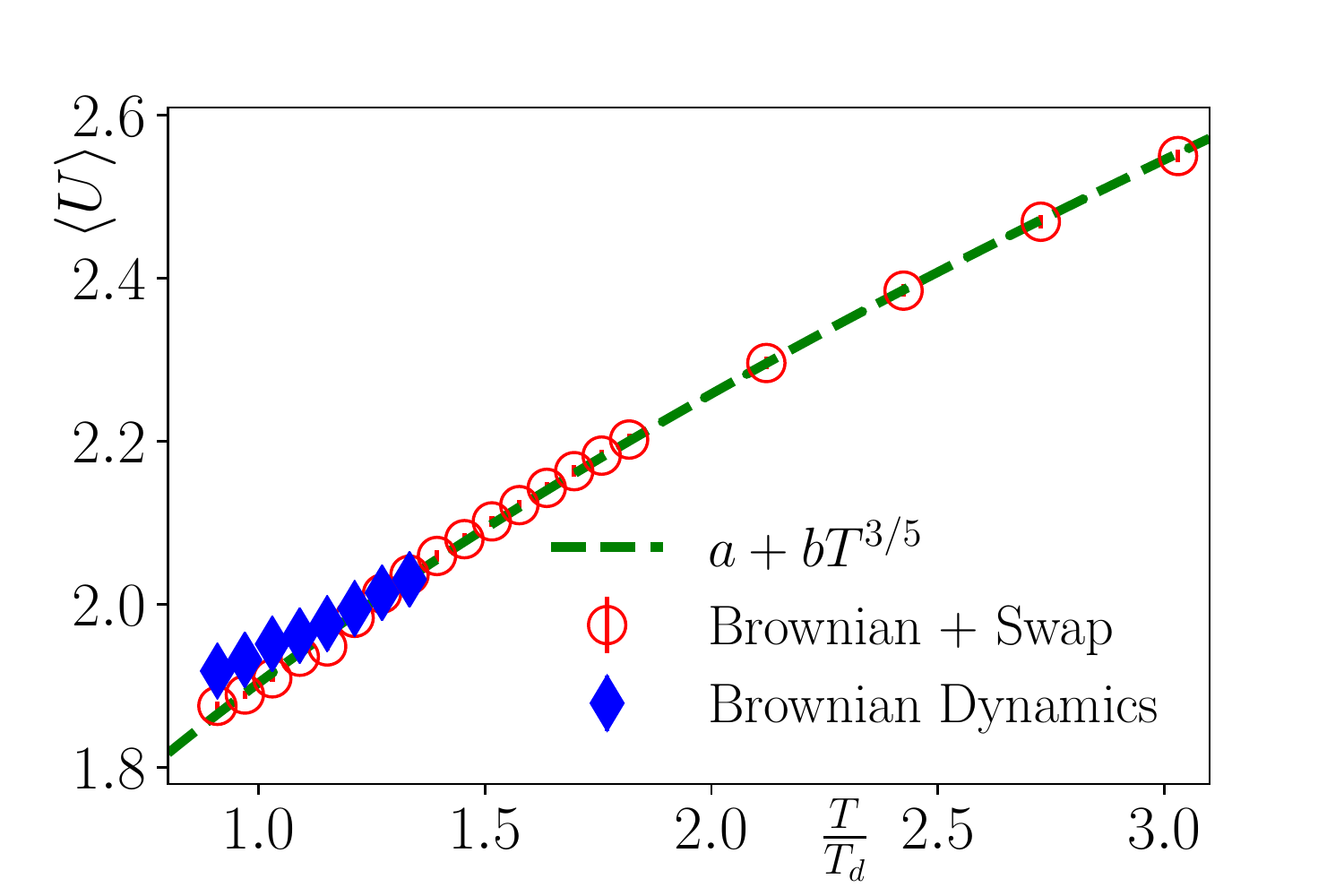}
\caption{(Colour online) Potential energy $\langle U \rangle$ as a function of temperature for system size $N=10^3$.
Blue diamonds refer to Brownian simulations, red circles refer to hybrid Brownian/Swap simulations. The dashed green line is the best fit to Rosenfeld and Tarazona formula \cite{rosenfeldTarazona1998density}.
$T_\text d$ is the dynamical temperature. \mttp{Blue symbols deviate from Rosenfeld and Tarazona formula indicating
that the corresponding stationary values of internal energy refer to configurations that are not well thermalized.}  
}\label{fig:energy}
\end{figure}

Figure \ref{fig:energy} reports the behavior of the internal energy $\langle U \rangle \equiv \langle N^{-1} \sum_{i<j} \phi(r_{ij} ) \rangle_{t}$, where
the angular brackets $\langle \dots \rangle_t$ indicate averages over trajectories, i.e., $\langle \mathcal{O} (t) \rangle_t \equiv \frac{1}{t_\text{fin}} \int_{t_0}^{t_0 + t_\text{fin}} \rd t\, \mathcal{O}(t) $, with
$\mathcal{O}(t)$ a generic observable, $t_0$ is chosen such that the starting configuration is equilibrated at the temperature $T$, and $t_\text{fin} \gg t_0$. Blue symbols refer to purely Brownian simulations, red symbols are hybrid Brownian/Swap
simulations. As one can see, data obtained through Brownia/Swap simulations are well fitted by Rosenfeld and Tarazona (RT) formula \cite{rosenfeldTarazona1998density}
indicating that they are well thermalized. On the contrary, blue symbols deviate from RT meaning that the corresponding configurations did not reach thermal equilibrium. 
The dynamical temperature of the model $T_\text d$ has been computed fitting 
the structural relaxation time 
$\tau_\alpha$ with a power law $\tau_\alpha \sim (T - T_\text d)^{-\delta}$.
$\tau_\alpha$ is defined as $Q(\tau_\alpha)=\re^{-1}$. $Q(t)$ is
the self-overlap function between two configurations of the system,
the first one taken at $t=0$ and the second one at $t$ \cite{Glotzer}. 
Dynamical quantities have been computed considering the Brownian evolution
of configurations that were previously thermalized through Brownian/Swap dynamics.

\subsection{Inherent Structures and Density of States}

\begin{figure}[!t]
\centering
\includegraphics[width=.5\textwidth]{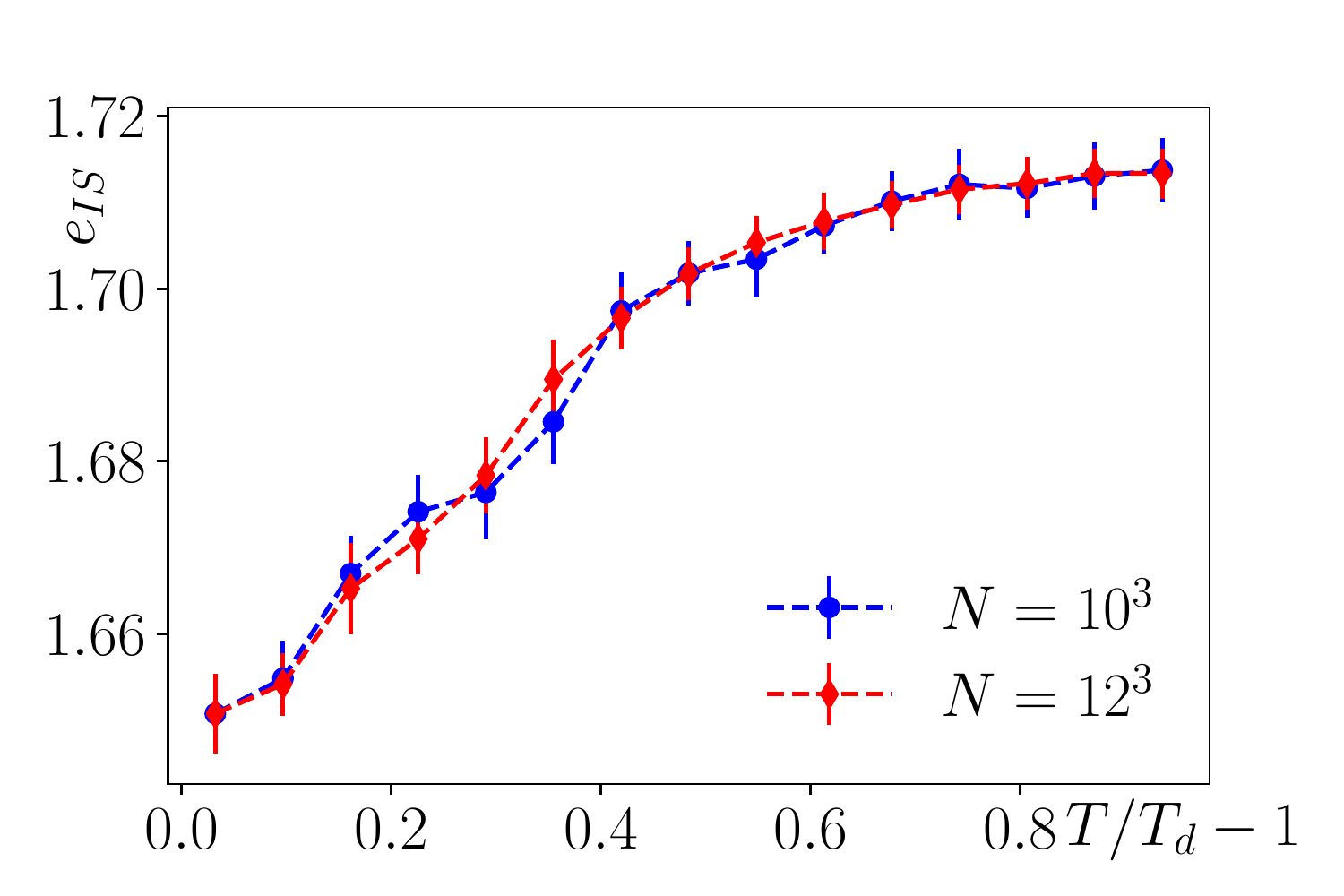}
\caption{(Colour online) Average energy of the Inherent Structures $e_\text{IS}$ as the parental 
temperature $T$ changes for two system size, i.e., $N=10^3$ (blue symbols) and $N=12^3$ (red symbols).  
}\label{fig:eis}
\end{figure}

After thermalization, we compute the corresponding inherent structures through
the Limited-memory Broyden-Fletcher-Goldfarb-Shanno
algorithm \cite{bonnans2006numerical}.
Let $\rr$ be a configuration of the system, i.e., $\rr \equiv (\rr_1,\ldots,\rr_N)$.
The mechanical energy of the configuration $\rr$ is $E[\rr]=\sum_{i<j}\phi(r_{ij})$. 
We indicate
with $\rr^0\equiv (\rr_1^0,\ldots,\rr_N^0)$ the configuration that minimizes $E[\rr]$,
We define the Inherent Structure energy $e_\text{IS} \equiv N^{-1} \langle E[\rr^0] \rangle_s$, with 
$\langle \dots \rangle_s$ indicating the average over $s$ independent configurations.
The \mttp{temperature dependence} of $e_\text{IS}$ is shown in figure \ref{fig:eis}.

The spectrum of the harmonic oscillations
around $\rr^0$ is then obtained considering a perturbed configuration $\rr = \rr^0 + \boldsymbol{\delta r}$. The mechanical energy now reads $E[\boldsymbol{\delta r}]=E[\rr^0] + \Delta E$
with $\Delta E \equiv \frac{1}{2} \sum_{i,j} \sum_{\mu \nu} \delta r_i^{\mu}M_{ij}^{\mu\nu}\delta r_j^\nu$ with $M_{ij}^{\mu\nu}$ the elements Hessian matrix $\mathbf{M}$, 
where Latin indices $i,j=1,\ldots,N$ indicate the particles and Greek symbols $\nu,\mu=1,\ldots,3$ the Cartesian coordinates.
To estimate the correlation length $\xi$, we also considered configurations where
a finite number of particles $pN$, with $p \in[0,1[$ the particle fraction, are maintained 
frozen during the minimization of $E[\rr]$. The details about the 
mi\-ni\-mi\-zation of $E[\rr]$ with pinned particles can be found in reference \cite{pnas}. 
We have computed all the $3N$ eigenvalues $\lambda_\kappa$, with $\kappa=1,\ldots,3 N$, 
using {\it gsl-GNU libraries} for sizes up to $N=12^3$. The corresponding {\textcolor{black}{eigenvalues}} 
of $\mathbf{M}$ are $\omega_\kappa^2=\lambda_\kappa$.

To identify the low-frequency spectrum, we focus our attention on the cumulative $F(\omega)$=$\int_0^\omega \rd \omega^\prime \, D(\omega^\prime)$, where $D(\omega)$=$\mathcal{N}^{-1} \sum_\kappa \delta(\omega - \omega_\kappa)$ is the density of states. $\mathcal{N}$ is the number of non-zero modes that is $3 N - 3$ for translational invariant systems.

The localization properties of the normal-modes have been investigated
through the inverse participation ratio 
${\textcolor{black}{\mathcal{R}}}(\omega)$ defined as  ${\textcolor{black}{\mathcal{R}}}(\omega) \equiv \sum_i | \ee_i (\omega )|^4  / \left( \sum_i | \ee_i (\omega )|^2 \right)^2
$ where $\ee_i(\omega)$ is the eigenvector of the mode $\omega$ \cite{Bell70}.
For a mode $\omega$ completely localized on a single particle, one has ${\textcolor{black}{\mathcal{R}}}(\omega)=1$, while a 
mode extended over all the particles corresponds to ${\textcolor{black}{\mathcal{R}}}(\omega)\sim N^{-1}$.
We also compute  $P(\mathcal{R})\equiv  \mathcal{Q} \langle  \sum_{\omega_\lambda : \omega_\lambda < \omega_\text{th}} \delta \left[ \mathcal{R}(\omega_\lambda) - \mathcal{R} \right] \rangle $
that is the probability distribution of \mttp{the inverse participation ratio for the modes with frequency}
$\omega$ smaller than a threshold frequency $\omega_\text{th}$. $\mathcal{Q}$ is a
normalization constant. After computing the distribution $P(\mathcal{R})$, we measure $\mathcal{R}_a \equiv \langle \log{\mathcal{R}} \rangle$.

\section{Results}

As it has been shown in reference \cite{pnas}, the non-Goldstone sector becomes clearly visible in $D(\omega)$, i.e., its weight in the
density of states overcomes phonons, in systems
with a finite fraction $p$ of frozen particles. In particular, non-Goldstone modes result
to be soft, i.e., $D(\omega) \sim \omega^{s(p)}$ with $s(p) > 2$ and localized, i.e., $\mathcal{R}(\omega)$ does not
scale as $1/N$. These facts are in agreement with the soft potential model
that predicts the scaling $D(\omega) \sim \omega^4$ for non-Goldstone excitations around the absolute minima in a one-dimensional
random energy landscape \cite{Gurarie03}.
Moreover, $D(\omega)$ at low frequencies changes its features when it is computed considering configuration\mttp{s} thermalized with
different protocols \cite{Lerner_rapid}. 
Thermalizing the system closer and closer to the
dynamical transition, the corresponding inherent structures show the same
type of crossover that can be described through a scaling $D(\omega)\sim\omega^{s(T)}$ \cite{arxiv}. Moreover, the same crossover can be documented 
through different observables, i.e., the effective exponent $s(T)$, the distribution of displacements
travelled by particles for reaching the inherent structures, the mean-distance travelled by a particle for reaching the optimal configuration. 
Here, we focus our attention on the $\mathcal{R}$ of the low-frequency modes
and their probability distribution function $P(\mathcal{R})$. We thus consider the inherent structures obtained  starting from configurations thermalized
at parental temperature $T$ that approaches the dynamical temperature $T_\text d$.

\subsection{Localization of the glassy modes}
\begin{figure}[!b]
	\centering
	\includegraphics[width=.5\textwidth]{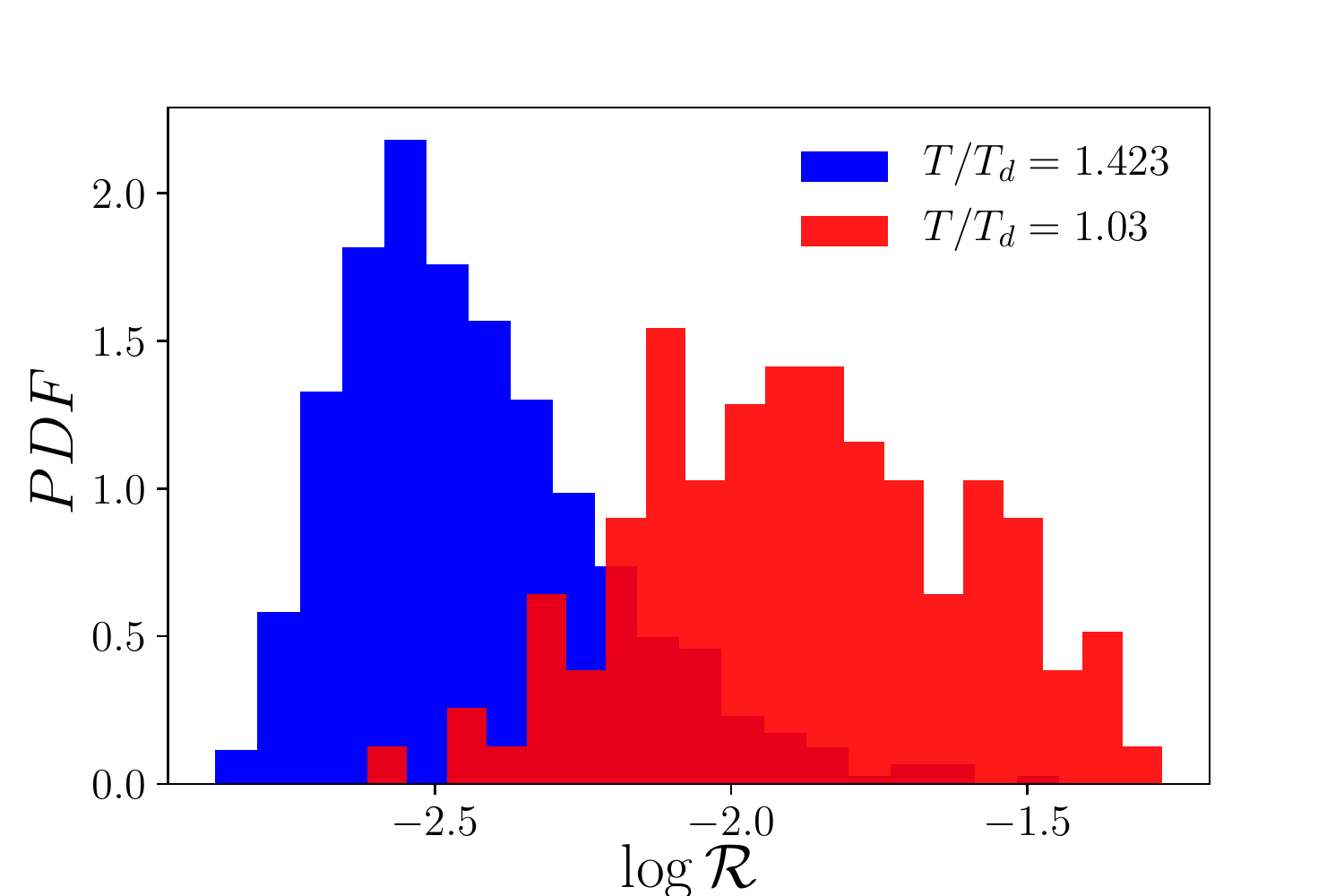} 
	\caption{(Colour online) Probability distribution $P(\mathcal{R})$ at temperatures $T/T_\text d=1.42$ (blue) and $T/T_\text d=1.03$ (red).  The cutoff frequency is $\omega_\text{th}=0.03$.
		\mttp{As temperature decreases, the distribution shifts at higher $\mathcal{R}$ values.}
	}\label{fig:IPR_temp}
\end{figure}

In figure \ref{fig:IPR_temp} we show the distribution  $P(\mathcal{R})$ at temperature $T/T_\text d=1.42$ (blue) and $T/T_\text d=1.03$ (red).
The distribution that has been computed considers only low-frequency modes. To do that, we impose a cutoff frequency $\omega_\text{th}$
that is chosen below the boson peak in the region where the power law scaling $D(\omega) \sim \omega^{s(T)}$ holds \cite{arxiv}. 
As one can appreciate, the distribution changes the shape and shifts towards higher $\mathcal{R}$ values
as temperature decreases indicating that extended modes have been progressively suppressed. 

\begin{figure}[!t]
	\centering
	\includegraphics[width=.5\textwidth]{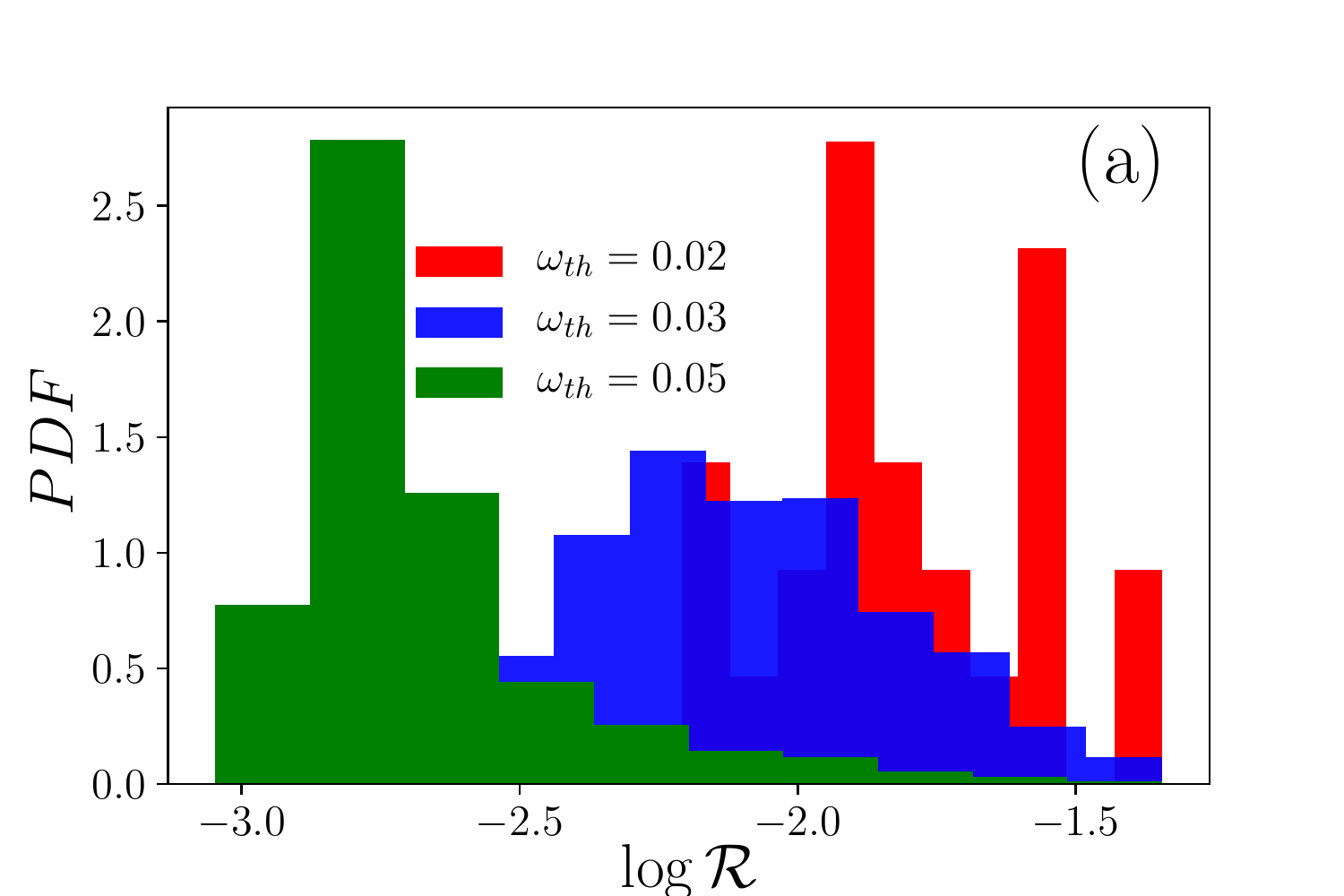} \\
	\includegraphics[width=.5\textwidth]{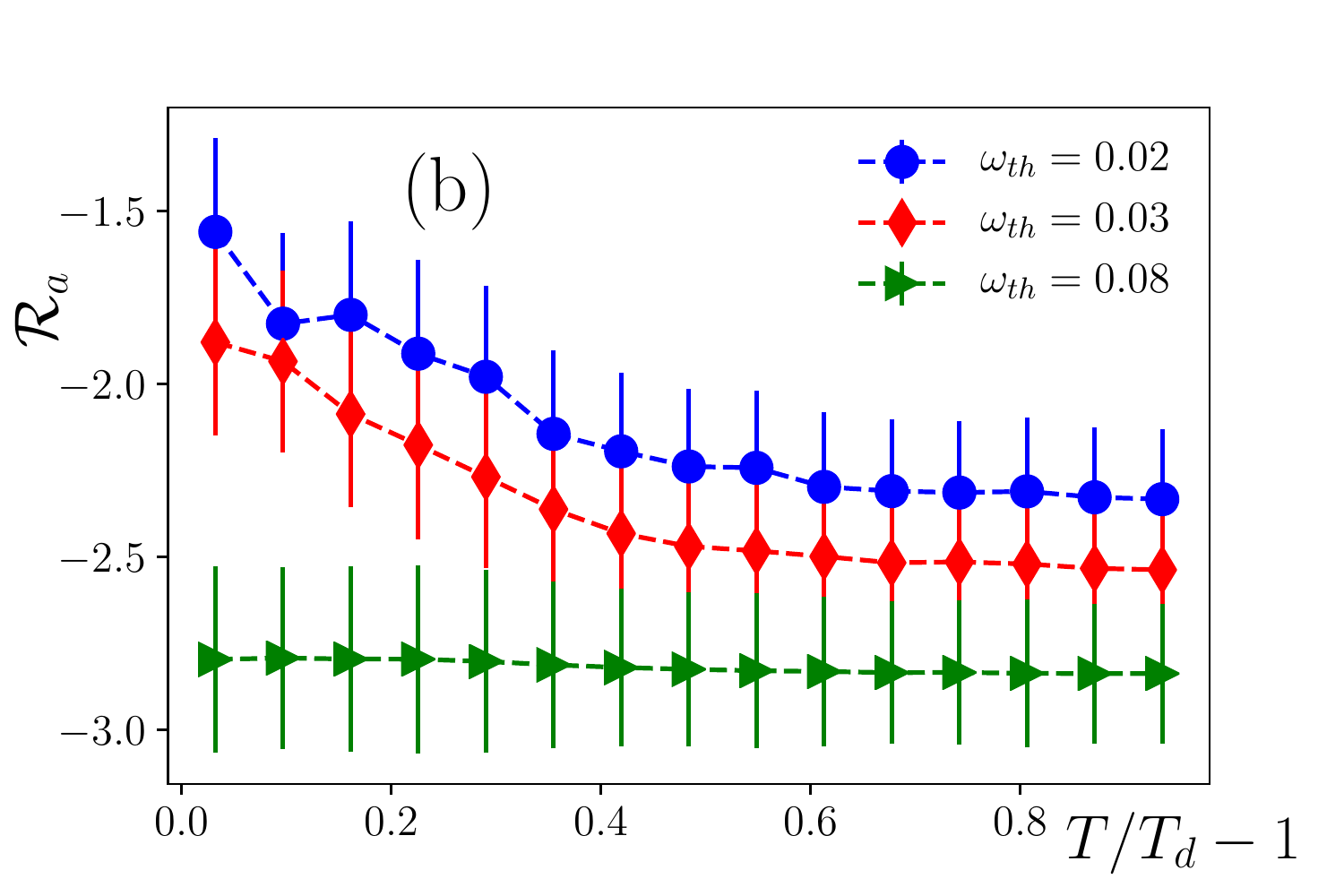}
	\caption{(Colour online) (a) $P(\mathcal{R})$ as $\omega_\text{th}$ increases from $0.02$ (red) to $0.05$ (green) \mttp{and $T/T_\text d-1=0.24$}. (b) $\mathcal{R}_a$ as a function 
		of the parental temperature for different choices of $\omega_\text{th}$.
	}\label{fig:IPR}
\end{figure}

To gain an insight into the role played by the cutoff frequency $\omega_\text{th}$, we
have computed $P(\mathcal{R})$  for different values of $\omega_\text{th}$. 
The cutoff $\omega_\text{th}$ is taken \mttp{below the frequency of the boson peak $\omega_\text{BP}$ which,
in our models, is around $\omega_\text{BP}\sim0.1$. 
The location of the boson peak in three spatial dimensions can be obtained looking at the maximum of 
$D(\omega) / \omega^2$ as shown, for instance, in reference \cite{grigera2003phonon}.}
\mttp{We thus choose $\omega_\text{th}\in[0.02,0.1]$.}
The result\mttp{s are} shown in figure \ref{fig:IPR}~(a) in the case of $T/T_\text d=1.16$.
The presence of extended modes  at frequencies larger than $0.04$ dramatically changes
the shape of  $P(\mathcal{R})$. This is made evident when one looks at $\mathcal{R}_a$ as a function of temperature as it is shown in figure~\ref{fig:IPR}~(b).   
\mttp{Choosing $\omega_\text{th}$ in the low-frequency region, i.e., $\omega_\text{th} < 0.08$, $\mathcal{R}_a$ undergoes a smooth crossover as
temperature decreases towards $T_\text d$.}
In particular, data for $\omega_\text{th}=0.02,0.03$ (blue symbols), i.e., a frequency that is
below the boson peak, show that $\mathcal{R}_a$ starts to increase for $T/T_\text d-1<0.5$. 


\begin{figure}[!t]
	\centering
	\includegraphics[width=.5\textwidth]{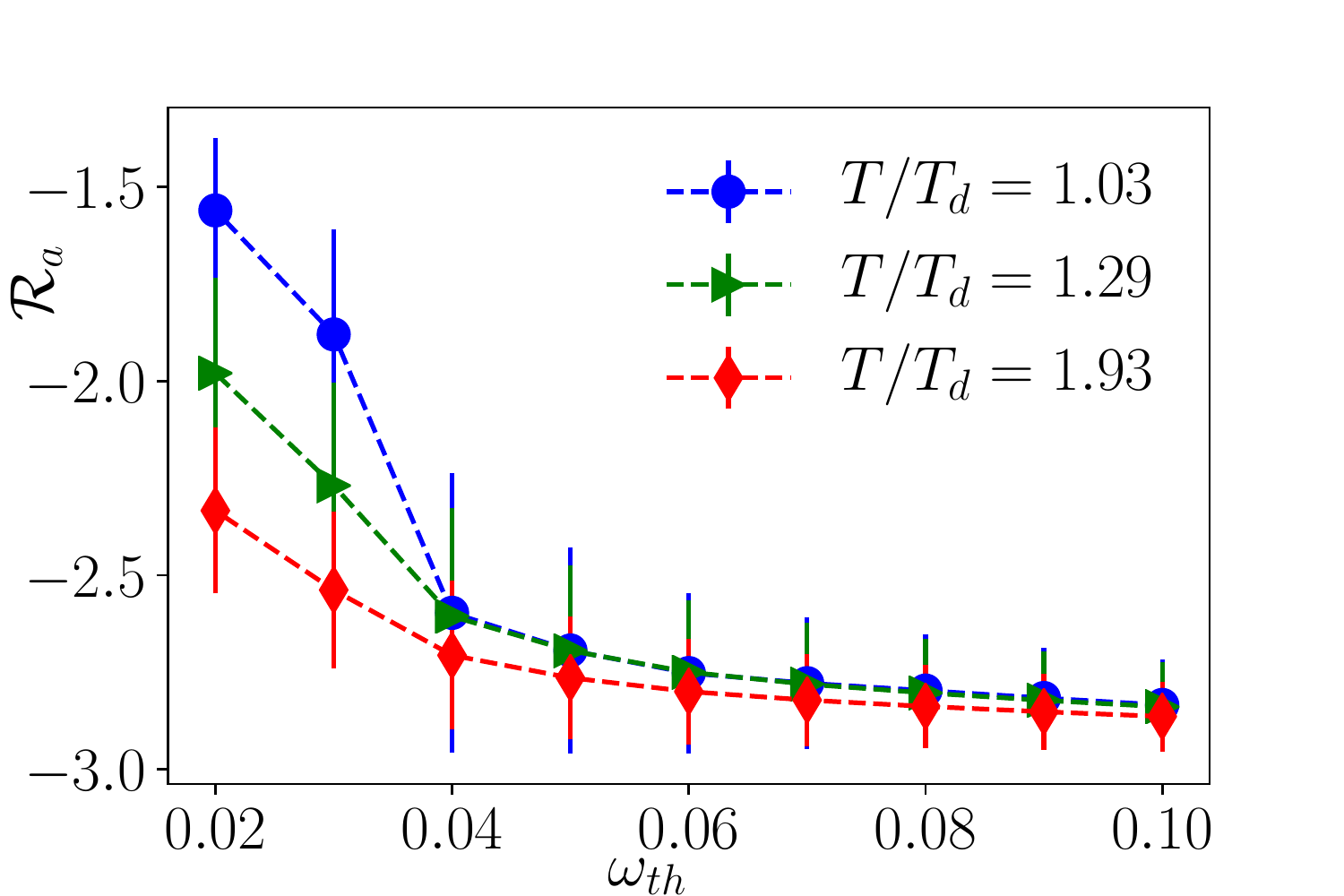} 
	\caption{(Colour online) $\mathcal{R}_a$ ad a function of the threshold frequency $\omega_\text{th}$ at increasing temperature values ($T/T_\text d=1.03,1.29,1.23$, red, green, and blue symbols, respectively). 
	}\label{fig:ipr_vs_th}
\end{figure}

\begin{figure}[!t]
	\centering
	\includegraphics[width=.5\textwidth]{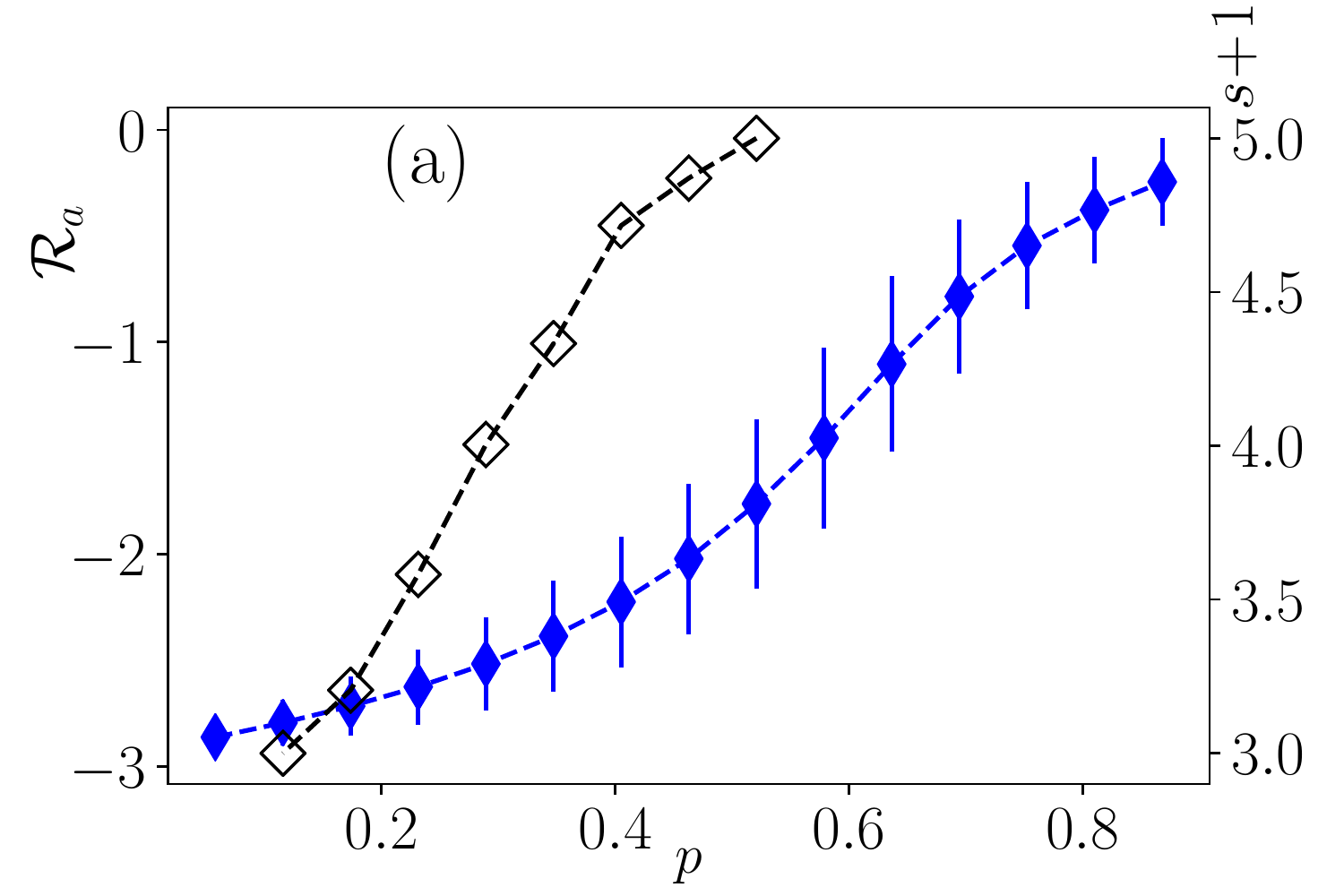} \\
	\includegraphics[width=.5\textwidth]{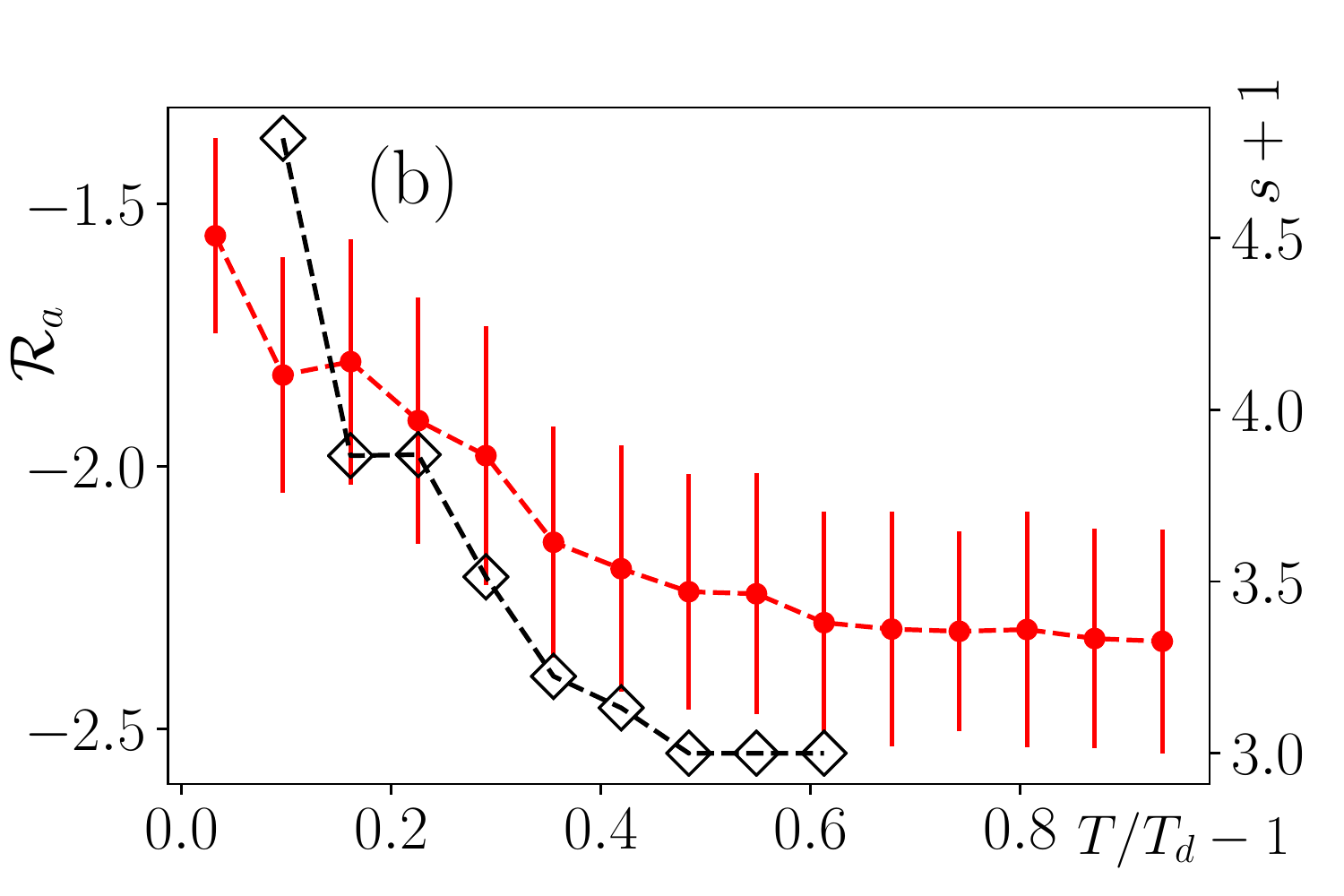}
	\caption{(Colour online) (a) $\mathcal{R}_a$ as a function of the fraction of pinned particles $p$ for configurations
		thermalized at high parental temperatures $T=0.5$. Open diamonds are
		the slope  $s(p)$ of the power law $D(\omega)\sim \omega^{s(p)}$.
		(b)~$\mathcal{R}_a$ as the parental 
		temperature decreases and $p=0$. Solid symbols are
		$\mathcal{R}_a$, open diamonds the slope $s(T)$ of the power law $D(\omega)\sim \omega^{s(T)}$.
	}\label{fig:IPR_comp}
\end{figure}

This finding is confirmed in figure~\ref{fig:ipr_vs_th} where we plot $\mathcal{R}_a$ as a function of $\omega_\text{th}$ for different
values of $T/T_\text d=1.03,1.29,1.93$ (blue, green, and red symbols, respectively). We observe a region that is independent of both, $\omega_\text{th}$
and parental temperature $T$. Below $\omega_\text{th} \sim 0.05$, $\mathcal{R}_a$ increases as $\omega_\text{th}$ decreases but also as $T$ decreases. This
is a clear signal that in the low-frequency region extended modes become attenuated in configurations thermalized at lower parental temperatures.
It is worth noting that the behavior of $\mathcal{R}_a$ as a function of $T$ mirrors the behavior of $s(T)$ as shown in figure\ref{fig:IPR_comp}~(b) where green symbols
are $s(T)+1$ obtained from the cumulative distribution $F(\omega)$ \cite{arxiv}.

\subsection{Comparison between $(T,p=0)$ and $(T=\infty,p)$ protocol}
Now we are going to compare the localization properties of the eigemodes obtained at high parental
temperatures but considering a finite fraction of a \mttp{frozen} particle during the minimization of the mechanical
energy \mttp{with those obtained in the non-pinned system as a function of $T$}. 
It has been shown that, as $p$ increases, $D(\omega) \sim \omega^{s(p)}$ with $s(p)$ undergoing 
a smooth crossover from $s=2$ to $s=4$ above a threshold value $p_\text{th} \sim 0.5$ \cite{pnas}. We refer to this 
protocol as $(T=\infty,p)$, since one considers configuration thermalized at high temperatures, i.e., away from $T_\text d$.
In the previous section, we have considered a protocol where the inherent structures were computed considering configutations
thermalized at parental temperatures $T$ close to $T_\text d$. We refer to this protocol as $(T,p=0)$ since energy minimization
has been computed without any artificially frozen particle.

Also in the case of the pinned system, modes responsible for $\omega^{4}$ are localized. This is because they live
in between the frozen regions induced by the random pinning protocol. 
Since the number of frozen
particle increases, the lowest frequency of the spectrum naturally shifts towards higher values and the 
threshold value $\omega_\text{th}$ as well. Random pinning explicitly destroy\mttp{s} the translational invariance removing the corresponding three zero
modes from the spectrum and attenuating any extended mode. This has a strong
effect on $\mathcal{R}_a$ that grows towards $\mathcal{R}_a \sim 0$ for $p\to1$, as it is shown in figure~\ref{fig:IPR_comp}~(a),
blue symbols. Also in this case, similarly to what we observed in the previous protocol, i.e., decreasing the parental temperature without any
artificially frozen particle, $\mathcal{R}_a$ undergoes a crossover mirroring that in $s(p)$ (black symbols in the same panel).
It is worth noting that in both protocols when the density of states undergoes a crossover $s \to 4$, $R_a \sim -1.5$.

\begin{figure}[!t]
	\centering
	\includegraphics[width=.5\textwidth]{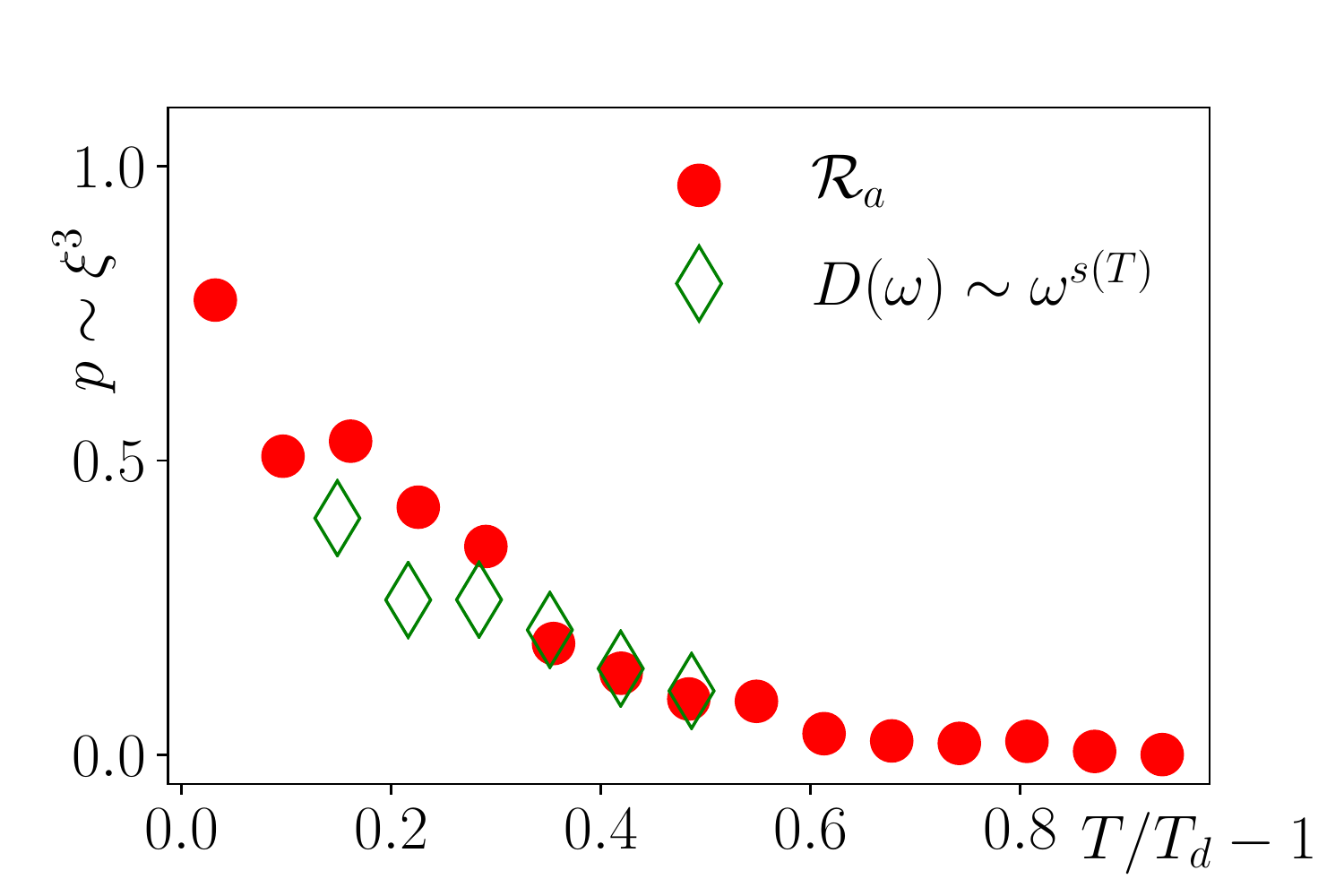} 
	\caption{(Colour online) Comparison between the typical length scale $\xi^3$ computed in reference \cite{arxiv} (open diamonds) with an estimate
		of $\xi^3$ based on
		$\mathcal{R}_a$ (solid circles). 
	}\label{fig:xi}
\end{figure}

Since frozen particles occupy a finite volume fraction $pV$, we can associate to the volume fraction a typical length 
scale $\xi \mttp{\equiv (p V)^{1/3}}$. In reference \cite{arxiv} it has been shown that, looking at the solution of $s(p,T=\infty)=s(T,p=0)$, one can
extract the behavior of $\xi^3$ as a function of the parental temperature $T$.  Here, we can extract $\xi^3$ looking at the
degree of localization of the low-frequency modes. To do that, we solve and invert numerically
$\mathcal{R}_a(T,p=0)
- \mathcal{R}_a(T=\infty,p=0)  = \mathcal{R}_a(T=\infty,p) -
\mathcal{R}_a(T=\infty,p=0.1)$. The result is shown in figure~\ref{fig:xi}, red symbols. Green symbols refer to the solution of $s(p,T=\infty)=s(T,p=0)$.  
\mttp{As one can see, the two data sets are in a good agreement. }

\section{Conclusions}
In this paper, we have investigated the localization properties of low-frequency modes in a three-dimensional
model of supercooled liquid. In particular, we have focused our attention on the role played by the parental temperature
on the localization of the soft glassy modes.
Our findings show that low-frequency vibrational modes at lower parental temperature turn out to
be more localized than those populating the density of states at higher $T$ values.
This is consistent with the
results presented in~\cite{arxiv,Lerner_rapid} and also with simulations on well-equilibrated polydisperse glass former \cite{wang2019low}.
In particular, the increasing in localization takes place near the dynamical transition temperature $T_\text d$ that is where the exponent $s(T)$
of the power law $D(\omega)\sim \omega^{s(T)}$ approaches $s(T_\text d)\to 4$. This finding confirms the interplay between dynamical
and zero-temperature structural properties in glasses \cite{arxiv}.
At lower temperatures, we  also observed a dependency of $\mathcal{R}_a$ on the cutoff frequency $\omega_\text{th}$. 
This shift reminds the effect of random
pinning on the density of states \cite{pnas}.

We have thus investigated how the localization of the lowest eigenmodes takes place in the same system with random pinning.
In particular, we observed that 
the same scenario of progressive localization of {\it glassy modes} \mttp{takes place} as the number of frozen particles increases.
In the pinning protocol, the emergence of soft localized excitations is due to the breaking of translational invariance in the system. 
With increasing $p$, moving particles rattle into small islands that are surrounded by the frozen ones. 
At higher values, \mttp{i.e., for $p > p_\text{th}$,} the phononic spectrum is totally destroyed giving rise to extremely localized modes, i.e., $\mathcal{R}\to1$.
This marks a difference with a system thermalized at lower parental temperatures where \mttp{the translational invariance is
preserved.}  \mttp{Nevertheless, configurations thermalized at lower temperatures show a spectrum of harmonic vibrations whose
properties are remarkably similar to those obtained breaking explicitly the spatial translational invariance, i. e.,
 crossover in $D(\omega)$ from Debye to non-Debye, localization of low-frequency modes, caging effects during minimization. 
 We can thus extract useful and complementary information comparing the region
$p < p_\text{th}$, in the $(T=\infty,p)$ protocol, with $T > T_\text d$, in the $(T,p=0)$ protocol.}
\mttp{ In this paper, we provide evidences for a crossover from extended to localized modes at low-frequencies as temperature
decreases towards $T_\text d$ which is in
agreement with recent works \cite{wang2019low,coslovich2018mode}. We also showed that,
in analogy with reference \cite{arxiv}, $\mathcal{R}_a$ can be employed for mapping  structural properties into dynamical ones. In particular,
the degree of localization measured through $\mathcal{R}_a$ is regulated by the proliferation of dynamical heterogeneous
regions in the $(T,p=0)$ protocol. The $(T=\infty,p)$ protocol allows one to define a length scale $\xi \sim p^{1/3}$ that
is an external and tunable parameter. We can thus study $p=p(T)$ just looking at the solution of $\mathcal{O}(T,p=0)=\mathcal{O}(T=\infty,p)$, with
$\mathcal{O}$ being a generic observable. As it has been shown in reference \cite{arxiv}, $p(T)$ mirrors the behavior of the
dynamical length scale $\xi_\text{dyn}$ \cite{Biroli2006}. Here, we showed that from $\mathcal{R}_a$ we can extract the behavior of 
$p(T)$ which is in agreement with those observed through other observables \cite{arxiv}.}

As a future direction, it would be interesting to study the density of states in systems thermalized at parental temperature 
$T$  close to $T_\text d$ with a fraction $p$ of pinned particles. In this way, in the spirit of early works on random pinning \cite{cammarota2013random_pin,cammarota2012ideal_pin,Szamel,kob2012non_pin,gokhale2014growing_pin,karmakar2013random,PhysRevLett.110.245702_pin,ozawa2015equilibrium_pin}, it would be possible to take access to the properties of {\it glassy modes} towards the
Kauzmann temperature.

\newpage
\section*{Acknowledgements} 
MP acknowledges the financial support of 
the Joint Laboratory on ``Advanced and Innovative Materials'', ADINMAT, WIS-Sapienza.

\newpage

\ukrainianpart

\title{Низькочастотні збудження та властивості їх локалізації у скловидних системах}
\author{М. Паолуцці\refaddr{adr1}, Л. Анджелані\refaddr{adr1,adr2}}
\addresses{
	\addr{adr1}Фізичний факультет, Римський університет ``Sapienza'', пл. A. Moро, 2, I-00185 Рим, Iталія
	\addr{adr2}ISC-CNR, Інститут складних систем, пл. A. Moро, 2, I-00185 Рим, Iталія	
}

\makeukrtitle

\begin{abstract}
	Крім сповільнення динаміки, що характеризується величезним ростом в'язкості при наближенні до переходу у
	скловидний стан, структурні стекла проявляють також цікаві аномальні термодинамічні особливості при низьких
	температурах, які вказують на специфічні відхилення від закону Дебая при достатньо низьких температурах. 
	Теорія, комп'ютерне моделювання, та експерименти припускають, що відхилення від закону Дебая є внаслідок 
	м'яких локалізованих скловидних мод, які заповнюють низькочастотний спектр. Ми досліджуємо властивості 
	локалізації низькочастотних мод у тримірній моделі переохолодженої рідини. Густина станів $D(\omega)$ 
	розраховується з розгляду властивих структур з конфігурацій, що були добре термалізовані при температурах близьких
	до динамічного переходу $T_\text d$. Ми спостерігаємо кросовер в розподілі ймовірностей для оберненого параметра внеску,
	що має місце при наближенні до $T_\text d$ з боку високих температур. Ми показуємо, що подібний кросовер спостерігається
	при високих температурах, коли трансляційна інваріантність системи явно порушується випадковим пінінговим полем.

	\keywords скловидні системи, динамічні властивості, густина коливних станів, комп'ютерне моделювання
	
\end{abstract}

\end{document}